\documentclass[a4paper,fleqn,usenatbib]{mnras}

\usepackage[T1]{fontenc}
\usepackage{ae,aecompl}

\usepackage{graphicx}	
\usepackage{amsmath}	
\usepackage{amssymb}	
\usepackage{rotating}

\newif\ifAMStwofonts                        
\newcommand{\lsimeq}{{_<\atop^{\sim}}}
\newcommand{\gsimeq}{{_>\atop^{\sim}}}

\title[Tension between SCUBA-2 and {\em Herschel} IR LFs]{On the existence of bright IR galaxies at $z$$>$2: tension between {\em Herschel} and SCUBA-2 results?}

\author[C. Gruppioni]{
Carlotta Gruppioni,$^{1}$\thanks{E-mail: carlotta.gruppioni@oabo.inaf.it}, Francesca Pozzi,$^{2,1}$
\\
$^{1}$Istituto Nazionale di Astrofisica -- Osservatorio di Astrofisica e Scienza dello Spazio, via Gobetti 93/3, I-40129, Bologna, Italy.\\
$^{2}$Dipartimento di Fisica e Astronomia, Universit\'a degli Studi di Bologna, via Gobetti 93/2, I-40129, Bologna, Italy.
}

\date{Accepted 2018 November 28. Received 2018 November 28; in original form 2018 July 30}

\pubyear{2018}

\begin{document}
\label{firstpage}
\pagerange{\pageref{firstpage}--\pageref{lastpage}}
\maketitle

\begin{abstract}
Recent derivations of the galaxy star formation rate density (SFRD) obtained from sub-millimetre (sub-mm)
surveys (e.g., SCUBA-2) show a tension with previous works based on {\em Herschel} and 
multi-wavelength data. Some of these works claim that the SFRD derived by pushing 
the {\em Herschel} surveys beyond $z$$\simeq$2 are incorrect. 
However, the current sub-mm surveys obtained from SCUBA-2 data and the methods used 
to construct the total infrared (IR) luminosity function (LF) and the SFRD could be affected by 
some limitations. 
Here we show how these limitations (i.e., selection bias and incompleteness effects) might 
affect the total IR LF, making the resulting dusty galaxy evolution of difficult interpretation.
In particular, we find that the assumed spectral energy distribution (SED) plays a crucial role
in the total IR LF derivation; moreover, we confirm that the long-wavelength (e.g., 850-$\mu$m) 
surveys can be incomplete against ``warm'' SED galaxies, and that the use of a wide spectral 
coverage of IR wavelengths is crucial to limit the uncertainties and biases. 

\end{abstract}




\section{Introduction}
To directly measure the star-formation (SF) activity in galaxies independently of any extinction corrections, one needs to observe in the far-infrared (far-IR) and sub-millimetre (sub-mm) domains. 
The amount of dust-obscured SF is directly linked to the measure of the total IR luminosity ($L_{\rm IR}$, integrated over the 8--1000 $\mu$m rest-frame range) 
through an empirical formula provided by \citet{kennicutt98} and widely used in the literature.
Therefore, how one derives the SF depends strongly on the shape of the assumed SED and on how well the far-IR bump (produced by dust, which re-emits in the IR the UV radiation from 
young and massive stars) is constrained. The peak and the shape of the far-IR bump provide important information about the dust temperature, the amount of dust and the galaxy type. 
For this reason, it is extremely important to obtain as many photometric data as possible, in order to constrain the galaxy SEDs over a wide range of wavelengths (possibly from UV to sub-mm/mm). 
As a consequence, to study how the star-formation rate density (SFRD) evolves with time across the Universe, it is necessary to constrain large samples of multi-wavelength galaxy 
SEDs over wide luminosity and redshift ranges. By counting these galaxies, per unit comoving volume, in luminosity and redshift bins we can derive the LF, one of the most 
important tools to study galaxy evolution.  
 
The extragalactic surveys performed with the {\em Herschel} observatory (e.g., PACS Evolutionary Probe, PEP, \citealt{lutz11}; Herschel Multi-tiered Extragalactic Survey, 
HerMES, \citealt{oliver12}; Herschel-GOODS, H-GOODS, \citealt{elbaz11}; Herschel-ATLAS, H-ATLAS, \citealt{eales10}) were the first ones to detect the peak of dust emission in 
galaxies up to high redshifts, and, due to the extensive multi-wavelength coverage in most of their fields, also to provide precise measurement of $L_{\rm IR}$ in thousands of 
galaxies spanning wide $L_{\rm IR}$, $z$ ranges. 
The deepest {\em Herschel} surveys performed with PACS (observing at 70, 100 and 160\,$\mu$m; \citealt{poglitsch10}) allowed us to measure $L_{\rm IR}$ and trace the SFRD evolution up 
to $z$$\sim$3--4 (i.e., \citealt{gruppioni13, magnelli13}), detecting unexpectedly large numbers of very bright sources (i.e., $L_{\rm IR}$$>$10$^{12}$ L$_{\odot}$) at $z$$\geq$2).
The presence of these IR-bright sources at high-$z$ was not expected from semi-analytic models (SAMs), which indeed largely under-predict the high SFRs observed in 
{\em Herschel} galaxies at $z$$\sim$2--3.  
Observations performed with the longer wavelength {\em Herschel} instrument, SPIRE (observing at 250, 350 and 500\,$\mu$m; \citealt{griffin10}), detected even
brighter IR galaxies and at higher $z$'s ($\gsimeq$6; e.g., \citealt{riechers13}, \citealt{lutz14}, \citealt{rowanrobinson16}, \citealt{laporte17}), but with large identification 
uncertainties due to possible source blending.
In fact, because of the large beam of the SPIRE instrument, i.e., FWHM$\simeq$18, 25 and 36 arcsec at 250, 350 and 500 $\mu$m respectively (see \citealt{swinyard10}), source blending 
and mis-identification can be critical issues for sources selected at these wavelengths.

Because of source blending, recent works deriving the total IR LF of galaxies from a 850-$\mu$m selection (e.g., \citealt{koprowski17} 
using SCUBA-2 on the James Clerk Maxwell Telescope, JCMT), claim that the high values of SFRD (or IR luminosity density, $\rho_{\rm IR}$) inferred 
from pushing the Herschel surveys beyond $z$$\simeq$2.5 are incorrect (e.g., \citealt{koprowski17}). Indeed, these works derive a much steeper 
bright-end of the total IR LF than the {\em Herschel} ones already at $z$$>$1, implying far fewer bright IR sources (and/or lower IR 
luminosities) and a smaller dust-obscured SFRD, at high-$z$. 
\citet{koprowski17} ascribes the cause of this discrepancy to the fact that the number and luminosity of $z$$\gsimeq$2 sources had been
severely overestimated by {\em Herschel} studies due to the high confusion and fraction of blended sources in SPIRE maps. 
Indeed, the sub-mm derived bright-end is not well constrained by data, especially at $z$$>$2, and the large discrepancy is mostly in luminosity 
ranges where only model extrapolations are reported (without any SCUBA-2 data).
The results of \citet{koprowski17} seem nevertheless in agreement with the Atacama Large Millimeter Array (ALMA) observations of the 
Hubble Ultra Deep Field by \citet{dunlop17} and with the results of stacking the 
deepest SCUBA-2 Cosmology Legacy Survey (S2CLS) images (\citealt{bourne17}), both finding a transition between obscured 
and unobscured SFRDs at $z$$\sim$3--4, with a steep high-$z$ decline following that shown by UV surveys (e.g., \citealt{bouwens15}).
However, we must note that the area covered by the ALMA survey is too small (e.g., $\sim$4.5 arcmin$^2$; \citealt{dunlop17}) to detect 
the most luminous objects shaping the bright-end of the LF. Moreover, the total IR LF is strongly sensitive to the SEDs considered for deriving $L_{\rm IR}$,
since integrating different SEDs over 8--1000 $\mu$m can result in $L_{\rm IR}$ differing by up to an order of magnitude, for a given 850-$\mu$m 
flux density.

If we cannot demonstrate without any doubt that all the bright IR sources detected by {\em Herschel} at $z$$>$1 are not blended, we can however show with 
very simple arguments that the sub-mm regime (i.e. $\geq$850 $\mu$m) is not optimally suited for selecting the complete samples of dusty 
sources (at least at $z$$<$4) needed to derive the total IR LFs. In fact, the 850-$\mu$m selection likely misses most of the "warmer" SED sources detected 
by {\em Herschel} and dominating the bright-end of the total IR LF (\citealt{gruppioni13}). 
Indeed even before the advent of {\em Herschel}, radio selected high-$z$ galaxies with far-IR luminosities comparable to those of sub-mm 
selected galaxies (SMGs) at similar redshifts, but not detected in the sub-mm, were found by \citet{chapman04} and associated to a population with hotter characteristic 
dust temperatures. Higher dust temperatures for these galaxies than similarly selected SMGs were then confirmed by {\em Herschel} observations (e.g., \citealt{chapman10, magnelli10}). 
Since even a small increase in the dust temperature implies a large decrease in observed sub-mm flux density, the shorter wavelengths of the far-IR emission peak cause 
the sub-mm waveband to potentially miss up to half of the most luminous, dusty galaxies at $z$$\sim$2 (\citealt{chapman04}).

The flux density limits of the S2CLS survey also play a 
significant role once converted to IR source completeness, with the $L_{\rm IR}$ range covered by sub-mm data being very narrow around the LF knee (i.e., $L^*$).

Finally, the SED choice and how well the peak of dust emission is constrained are crucial. 
All these factors conspire in making the total IR LFs and SFRD results from 850-$\mu$m surveys rather uncertain or incomplete.

Throughout the paper, we use a \citet{chabrier03} stellar initial mass function (IMF) and adopt a $\Lambda$CDM cosmology with 
$H_{\rm 0}$\,=\,70~km~s$^{-1}$\,Mpc$^{-1}$, $\Omega_{\rm m}$\,=\,0.3, and $\Omega_{\rm \Lambda}\,=\,0.7$. 

\section{The main ingredients of the total IR LF and SFRD}
\label{wrong_ingred}
\subsection{SED as key building block of L$_{\rm IR}$}
\label{sed}
\begin{figure}
	\includegraphics[width=\columnwidth,height=7.2cm]{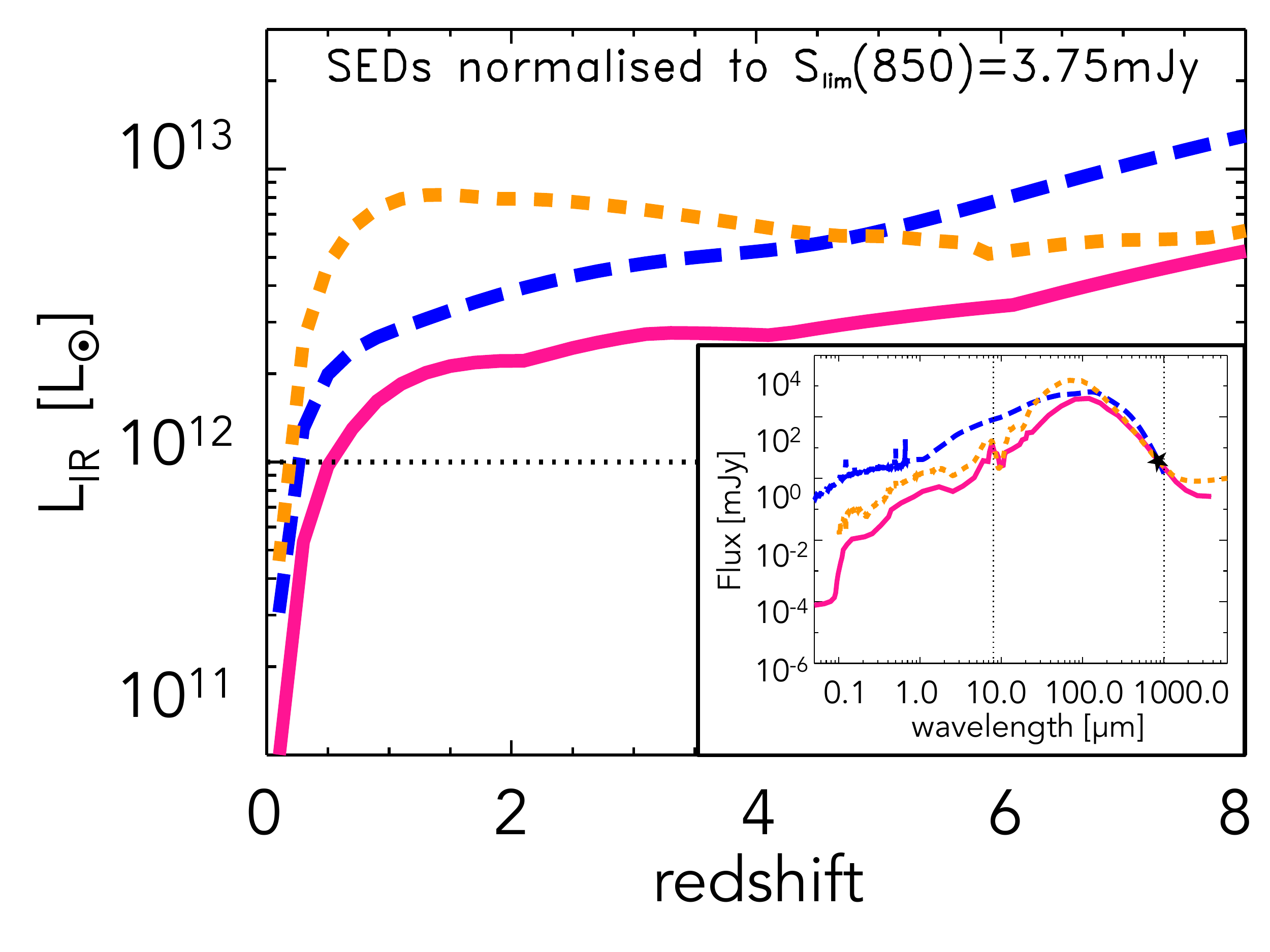}
\caption{$L_{\rm IR}$ obtained by integrating three different SEDs (two SEDs representative of the bulk of the {\em Herschel} population at $z$$>$2: IRAS20551, orange dashed, and a QSO, blue 
long-dashed, see \citealt{gruppioni13}), and the average sub-mm SED derived by \citealt{michalowski10}, deep-pink solid) normalised to $S$(850)$=$3.75 mJy, as a function of $z$.
    In the insert the three SEDs are shown (normalised to the same 850-$\mu$m flux). 
    }   
     \label{fig:LIR_SED}
\end{figure}
First, we discuss and analyse the key role played by the choice of galaxy SED in the $L_{\rm IR}$ derivation. The better the far-IR bump is constrained by data,
the more reliable the derived $L_{\rm IR}$ are. In fact, if the far-IR bump is not sampled by data points, one could obtain very different values of the 8--1000 $\mu$m luminosity
for the same 850-$\mu$m flux, depending on the considered SEDs. 
To derive the total IR LFs from {\em Herschel} data, \citet{gruppioni13} fitted each object with a set of different templates and
used the best-fit SED to compute the total IR LF: this provided the more realistic measure of the total IR luminosity source-by-source.
On the contrary, the multi-wavelength data relative to the S2CLS surveys of \citet{koprowski17} were used only
for estimating the photometric redshifts, while $L_{\rm IR}$ was computed from the observed 850-$\mu$m flux density
by just assuming for all the sources the average sub-mm galaxy template derived by \citet{michalowski10}, regardless of type, redshift or luminosity. 

In Figure~\ref{fig:LIR_SED} we show the variation of $L_{\rm IR}$ with $z$, obtained by integrating three different SEDs (normalised to the same 850-$\mu$m flux density, 
corresponding to the limiting flux of the SCUBA-2 Ultra Deep Survey, UDS, by \citealt{koprowski17}: $S$(850)$=$3.75 mJy): the average sub-mm SED by \citet{michalowski10} considered 
for all the SCUBA-2 sources (deep-pink solid), IRAS 20551 (orange dashed) and a typical QSO SED (blue long-dashed). The latter templates are those that best reproduce the SEDs of the bulk 
of {\em Herschel} sources responsible for the observed bright-end of the total IR LF at $z$$>$2 (\citealt{gruppioni13}). The three templates are shown in the insert at the
bottom right corner of the Figure.
It is clear that by integrating the three templates over the 8--1000 $\mu$m range
one would get very different results, even starting from the same 850-$\mu$m flux: it is therefore crucial to have data in the rest mid-/far-IR range to better constrain the SEDs
and $L_{\rm IR}$. 
Up to at least $z$$\simeq$8 the two best-fit {\em Herschel} SEDs provide much higher values of $L_{\rm IR}$ than the \citet{michalowski10} template. 
It is therefore not surprising that the SCUBA-2-based total IR LF has such a steep bright-end (see Figures 7 and 8 of \citealt{koprowski17}): $L_{\rm IR}$ 
can be severely underestimated (up to a factor of 5 at $z$$\simeq$1--2) causing the luminous objects to fall in the wrong (e.g., lower) luminosity bin. 
In order to quantify the effect of the SED, we have recomputed the total IR LF with $L_{\rm IR}$ derived from the \citet{michalowski10} template 
(normalised to the measured 160-$\mu$m flux) for all our PEP sources: the resulting LFs show a different shape in any $z$-bins (i.e., steeper than the \citealt{gruppioni13}, 
with lower or absent data points in the brighter L-bins and higher in the faint L-bin). The effect is more evident at lower $z$, while at higher redshifts the difference is 
less pronounced, due to the fact that the 850-$\mu$m band with increasing $z$ samples rest-frame wavelengths closer and closer to the SED peak, 
although the steepening of the LF is still significant.
In Table~\ref{tab_Phi_SED} we report the ratio between the two IR LFs calculated with the \citet{michalowski10} single SED and with the template library used by \citet{gruppioni13},
in two redshift bins (i.e., at 0.0$<$$z$$<$0.3 and 2.5$<$$z$$<$3.0).
\begin{table}
 \caption{Total IR LF ratio (single SED vs. SED library)}
\begin{tabular}{lcc}
\hline \hline
$<log_{10}(L_{\rm IR}/L_{\odot})>$     &  \multicolumn{2}{c}{$\Phi_{\rm michalowskiSED} / \Phi_{\rm gruppioniSEDs}$} \\
                                             &  0.0$<$$z$$<$0.3 & 2.5$<$$z$$<$3.0 \\ \hline
   8.2                                    &   4.3    & -- \\
    8.7                                    &   2.1   & -- \\
    9.2                                    &   0.5 & -- \\
    9.7                                   &   0.4  & -- \\
    10.2                                   &   0.2  & -- \\
     10.7                                   &   0.2  &  --\\
     11.2                                   &   0.02  &  --  \\
        11.7                                   &   0.0  &  --  \\
           12.2                                   &   0.0  &   4.6 \\
           12.7                                 &  --   &   0.9  \\
           13.2                                  & --   &   0.3   \\ 
           13.7                                   & --  &    0.1   \\   \hline \hline
           \end{tabular}
 \label{tab_Phi_SED}
           \end{table}
                      
\subsection{Sample selection and completeness}
\label{completeness}
\begin{figure}
	\includegraphics[width=\columnwidth,height=6.9cm]{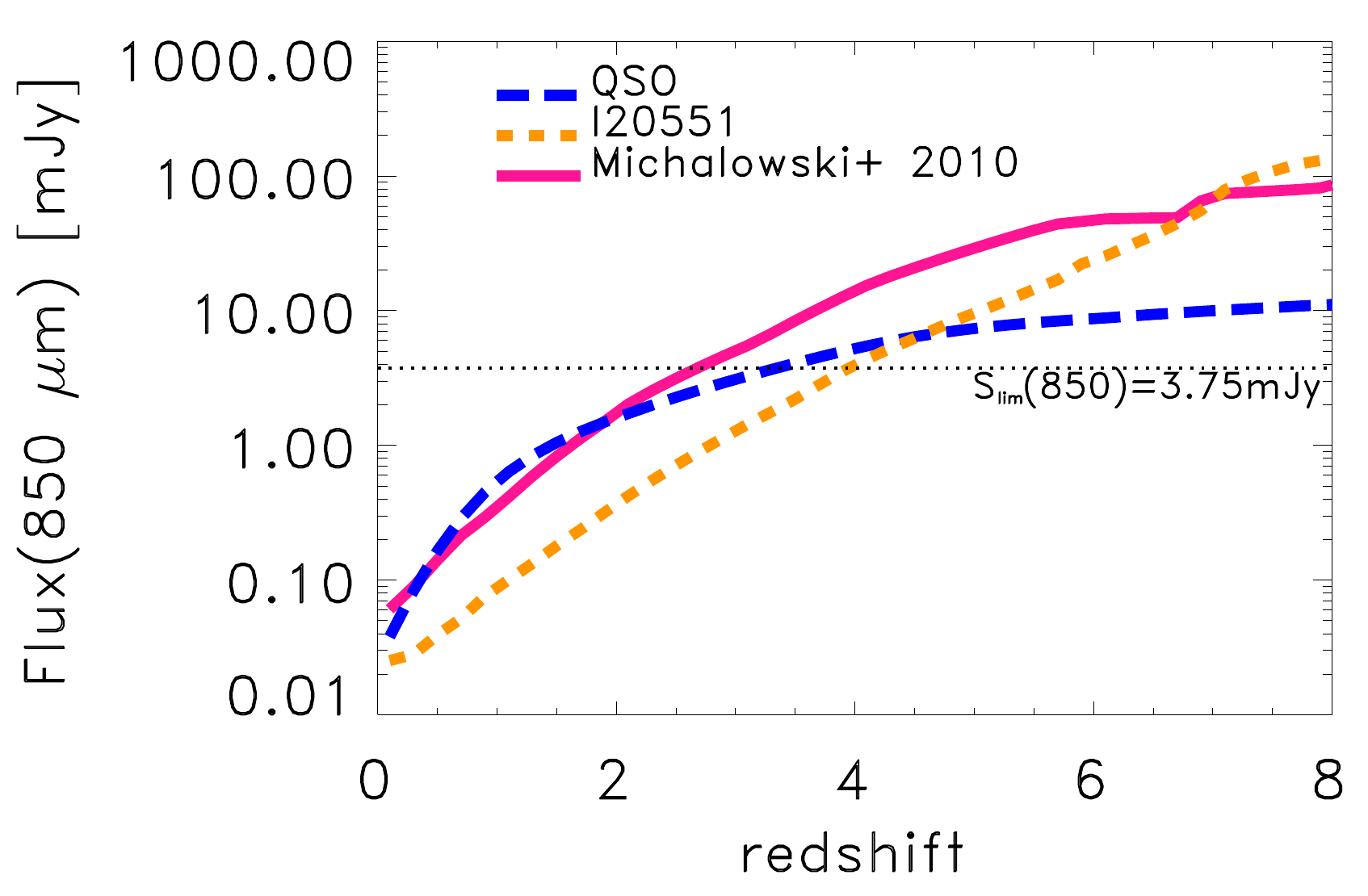}   
	\caption{Expected 850-$\mu$m flux for a {\em Herschel} source detected at the limiting flux of the deepest PEP survey (GOODS-S, $S$(160~$\mu$m)=2.4 mJy), 
	based on the three different SEDs described in
    figure~\ref{fig:LIR_SED}. The horizontal dotted line shows the limiting flux of the 850-$\mu$m SCUBA-2 UDS survey. }
    \label{fig:850flux}
\end{figure}
\begin{figure}
	\includegraphics[width=\columnwidth]{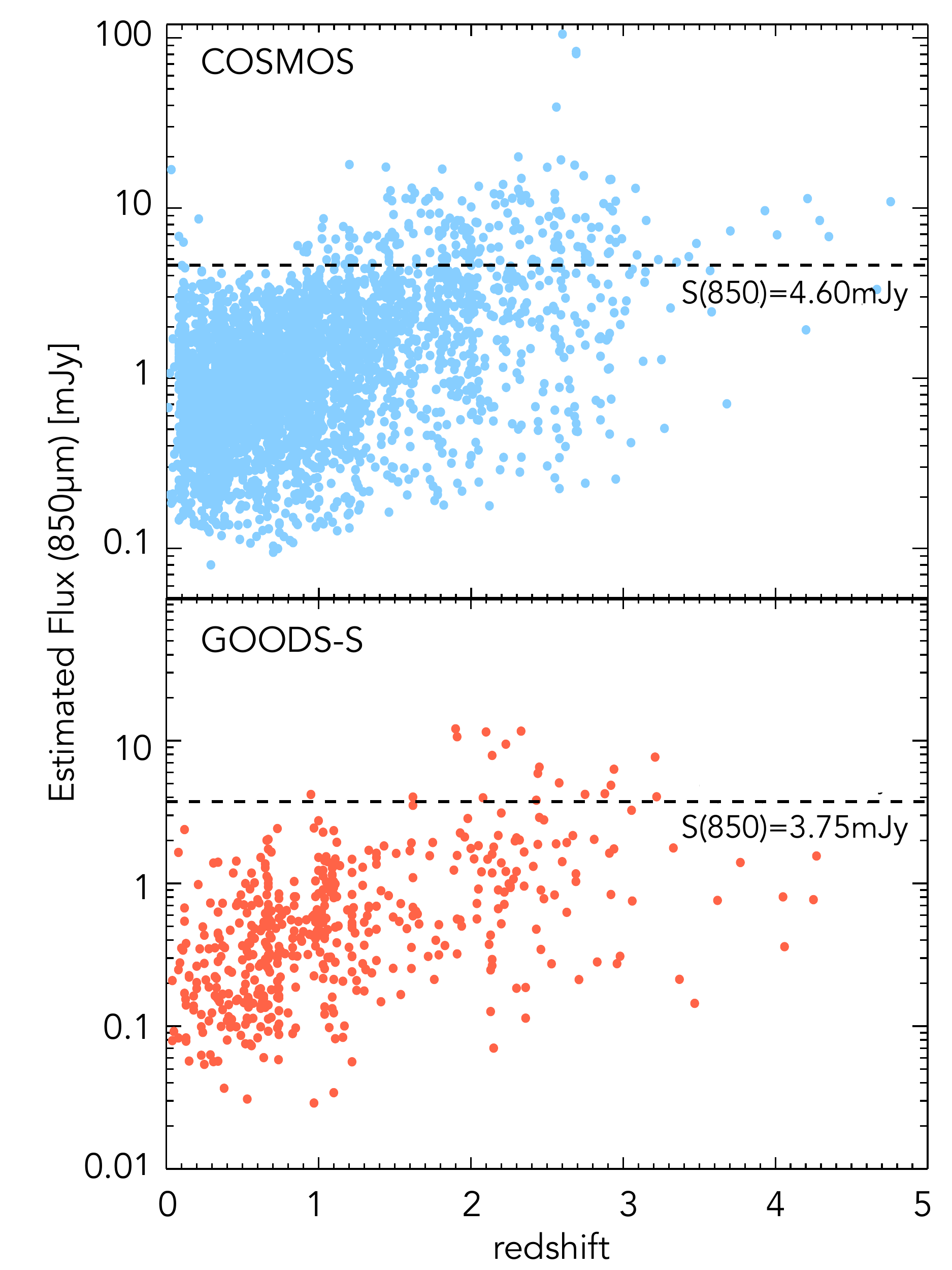}
    \caption{Flux at 850 $\mu$m expected for the {\em Herschel}-PEP 160-$\mu$m sources detected in the COSMOS ($top$, light-blue filled circles) and
    GOODS-S ($bottom$, orange-red circles) fields (based on their best-fit SEDs). The horizontal dotted lines show the 850-$\mu$m limits of the S2CLS 
    surveys in the COSMOS and UDS fields.}
    \label{fig:PEPflux850}
\end{figure}

Another factor that might conspire to depress the bright-end of the \citet{koprowski17} LF is the IR source incompleteness due to
the sub-mm selection. As an example, in Figure~\ref{fig:850flux} we show the 850-$\mu$m flux expected for the three template SEDs considered above,
for a source detected at 160 $\mu$m at the limiting flux of the PEP GOODS-S survey ($S$(160)$=$2.4 mJy). A source with the \citet{michalowski10}
SED (deep-pink), selected at the limits of {\em Herschel}-PEP, will not be detectable in the S2CLS UDS at $z$$<$3, while it will not be detectable up to 
$z$$\simeq$3.5--4 if it had a QSO- or IRAS 20551-like SED. 
Since the IR sources detected by {\em Herschel} show a wide diversity of SEDs, fitted by different templates, and the templates that best reproduce the observed
SEDs of most of the $z$$\gsimeq$2 {\em Herschel} galaxies making up the controversial bright-end of the IR LF are significantly ``warmer'' than the \citet{michalowski10} one 
(see, e.g., Figs. 1, 18 of \citealt{gruppioni13}),
we conclude that a 850-$\mu$m survey (unless extremely deep, e.g., to  $<$0.1 mJy) is, by definition, unable to detect the majority of the
{\em Herschel} sources at redshifts lower than $z$$\simeq$3--4. 
As an empirical probe of the above assertion, in Figure~\ref{fig:PEPflux850} we show the expected 850-$\mu$m flux density for all the {\em Herschel}-PEP 160-$\mu$m sources in
the COSMOS ($top$) and GOODS-S ($bottom$) fields contributing to the total IR LF and SFRD (\citealt{gruppioni13,gruppioni15}):
each flux density has been obtained by interpolating the best-fit template SED (reproducing data from UV to far-IR) for each source, and has to be considered only indicative 
(extrapolation from the longer wavelength sampled by {\em Herschel} to 850-$\mu$m can be very uncertain, being in the steep Rayleigh-Jeans domain).
Most of the {\em Herschel}-PEP sources are expected to be undetectable in the SCUBA-2 surveys of \citet{koprowski17}, and indeed from a cross-match between
the PEP and the S2CLS catalogues we find that only $\sim$5 per cent of the PEP 160-$\mu$m sources considered for the LF in the COSMOS field have a counterpart 
in the S2CLS catalogue (this percentage reduces to 2 (1) per cent if we limit to $z$$<$2 (1)).
We therefore cannot consider the total IR LF and SFRD derived from the S2CLS surveys as a complete derivation, since a population of bright 
and ``warm'' IR sources significantly contributing to these quantities is likely missed by them. 
\begin{figure}
	\includegraphics[width=\columnwidth,height=6.9cm]{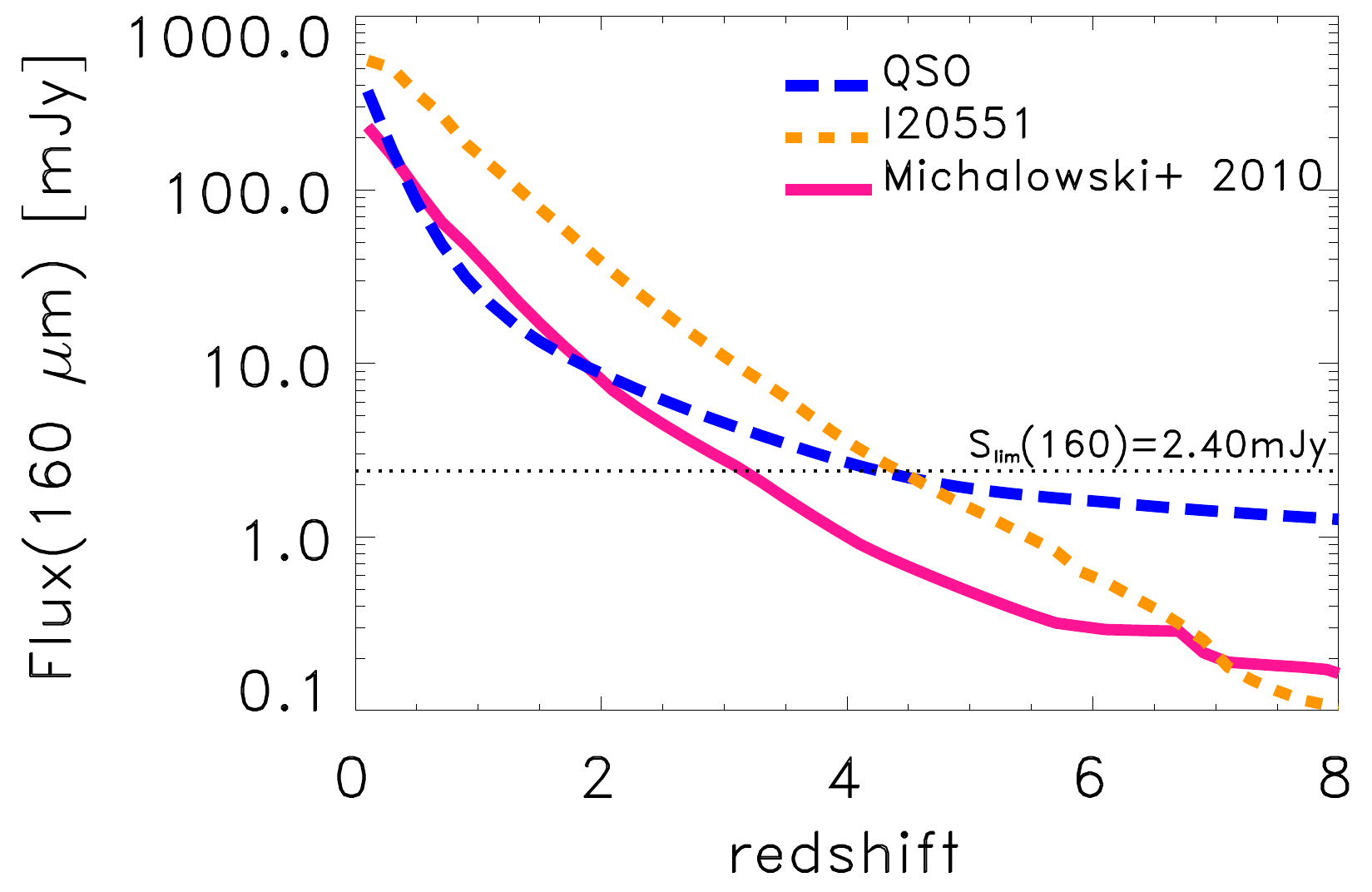} 
\caption{Expected 160-$\mu$m flux of a SCUBA-2 source with $S$(850~$\mu$m)=3.75 mJy, based on the SEDs of Figure~\ref{fig:850flux}. The horizontal dotted line shows the 
    limiting flux of the deepest 160-$\mu$m PEP survey in the GOODS-S.}
    \label{fig:160flux}
\end{figure}

The reverse case of the expected detectability of a S2CLS source in the PEP survey (e.g. in the GOODS-S field)
is plotted in Figure~\ref{fig:160flux}: depending on the SEDs, the SCUBA-2 sources are expected to be detected above the PEP flux density
limit at $z$$\lsimeq$3 ($\lsimeq$4 for an IRAS~20551 or QSO template). We therefore expect the deepest PEP 160-$\mu$m survey to be almost complete 
with respect to the faintest S2CLS sources, at least up to $z$$\simeq$3--4. Indeed, from a simple catalogue-catalogue cross-match within the same fields and in the overlapping areas, 
we found that $\sim$85\% of the S2CLS sources detected at S/N$>$4$\sigma$ have a PEP 160-$\mu$m counterpart in the GOODS-N field and $\sim$80\% in 
COSMOS (note that the PEP and S2CLS surveys in these two fields are shallower than those considered in Figure~\ref{fig:160flux}, that reach the deepest PEP and S2CLS
limits, but are not on the same area; moreover no redshift information was given for the S2CLS sources, so we are not able to test the percentage of counterparts as a function of $z$).

\subsection{Multiple sources}
\label{multiple}
The main reasons ascribed by \citet{koprowski17} to the large discrepancy between their and {\em Herschel} LFs are ``problems in source identification and redshift estimation arising
from the large-beam long-wavelength SPIRE data, as well as potential blending issues''. Source blending (or confusion) means that more than one astronomical source may be present 
within the beam, hence some sources can have their fluxes boosted. 
We have tested the effect of possible blending of SPIRE sources in our calculations by recomputing $L_{\rm IR}$ and the total IR LF by halving the 250, 350 and 500-$\mu$m fluxes
(i.e., in the extreme assumption that all the SPIRE fluxes come from the blend of two sources) and re-fitting the SEDs. The comparison between the new total IR LFs and the original ones indicates
a slight steepening when reducing the SPIRE fluxes, though not as extreme as found by using the \citet{michalowski10} template for all the sources (see Section~\ref{wrong_ingred}).
In Table~\ref{tab_Phi_blend} we report the ratios between the two IR LFs in two redshift bins (i.e., at 0.0$<$$z$$<$0.3 and 2.5$<$$z$$<$3.0).
\begin{table}
 \caption{Total IR LF ratio (f$_{SPIRE}$$\times$0.5 vs. original)}
\begin{tabular}{lcc}
\hline \hline
$<log_{10}(L_{\rm IR}/L_{\odot})>$     &  \multicolumn{2}{c}{$\Phi_{\rm michalowskiSED} / \Phi_{\rm gruppioniSEDs}$} \\
                                             &  0.0$<$$z$$<$0.3 & 2.5$<$$z$$<$3.0 \\ \hline
    8.7                                    &   2.4   & -- \\
    9.2                                    &   1.1 & -- \\
    9.7                                   &   1.1  & -- \\
    10.2                                   &   1.0  & -- \\
     10.7                                   &   0.7  &  --\\
     11.2                                   &   1.0  &  --  \\
        11.7                                   &   1.0  &  --  \\
           12.2                                   &   0.9  &   2.0 \\
           12.7                                 &  --   &   0.9  \\
           13.2                                  & --   &   0.6   \\ 
           13.7                                   & --  &    0.4   \\   \hline \hline
           \end{tabular}
 \label{tab_Phi_blend}
           \end{table}
Therefore, we can conclude that, even if all the SPIRE fluxes would result from the blend of two sources (e.g. a factor of 2 higher than the real ones), 
the error on the calculation of the total IR LF would be smaller by constraining the SEDs with all the other available data, than by using an average 
template for all the sources neglecting the observed data. 
However, these ratios are much lower than the differences claimed by \citet{koprowski17} at high $L_{\rm IR}$ (i.e., factors $>$100 at $z$$>$1.2--1.3),
that would imply that the SPIRE bright fluxes were likely due to the blend of more than two sources (though without a 24-$\mu$m counterpart):
these fainter sources artificially boosting the bright end of the {\em Herschel} LF by such large amounts, if resolved by SCUBA-2, should then produce a
steeper faint-end in the SCUBA-2 total IR LF (i.e., steeper than found by just halving the SPIRE fluxes), to compensate the much lower bright-end. 
At odds with this expectation, \cite{koprowski17} find a faint-end of the total IR LF in fairly good agreement with \citet{gruppioni13} and significantly 
shallower than the \citet{magnelli13} one.
\begin{figure*}
	\includegraphics[width=16cm,height=12cm]{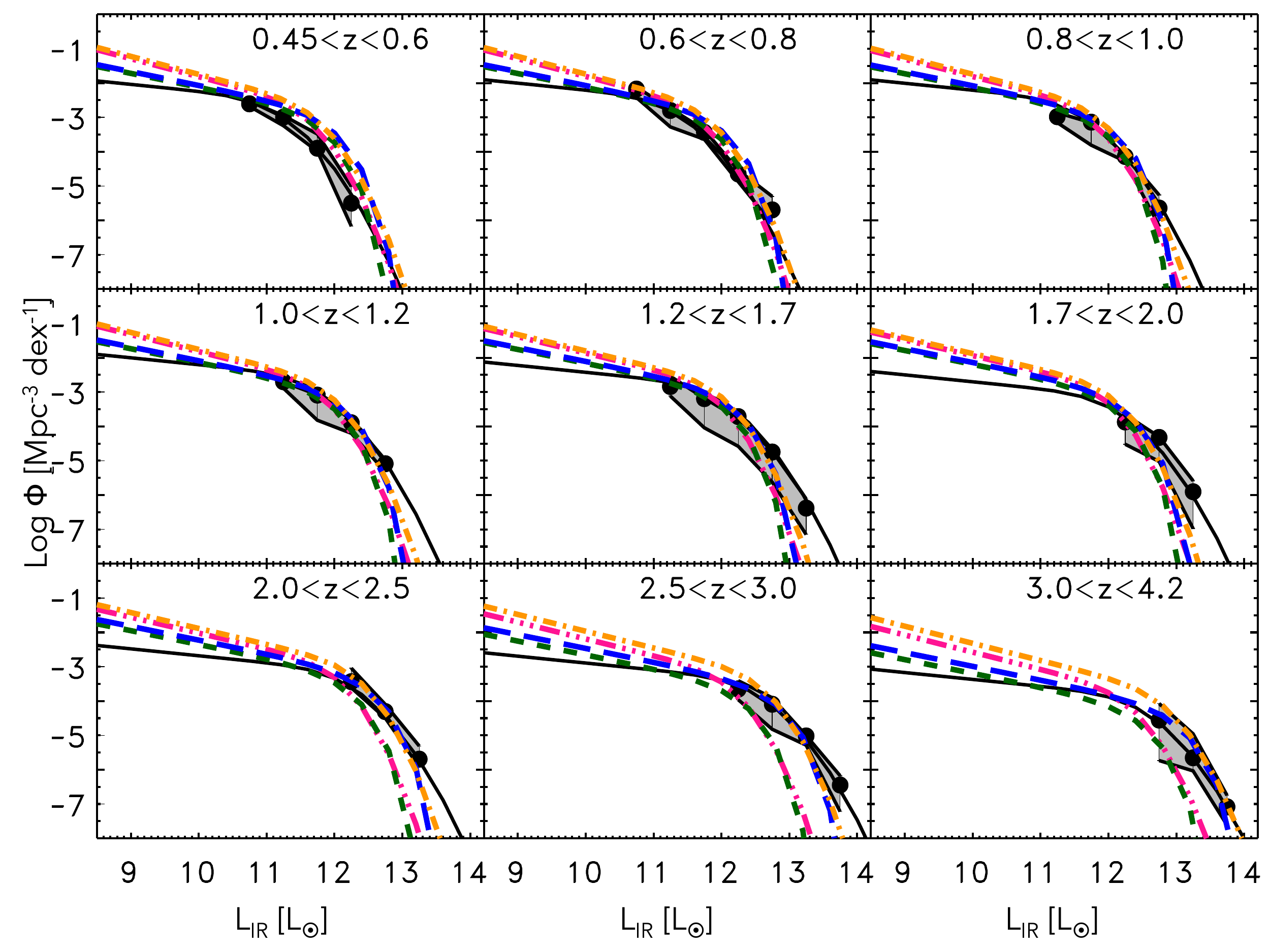}
    \caption{Comparison between the {\em Herschel}-PEP total IR LF by \citet{gruppioni13}, shown by the black filled circles within the grey
filled area, and different functional forms best-fitting the rest-frame 250-$\mu$m LF from SCUBA-2 data derived by \citet{koprowski17}. The dark-green dashed line shows the
    original Schechter function converted to total IR LF by using the $L_{\rm 250\mu m}$/$L_{\rm IR}$ ratio given by the 
\citet{michalowski10} template, then interpolated in $z$ to match the same redshift bins. The deep-pink dot-dot-dot-dashed line shows the result of the same procedure for a modified-Schechter 
function (e.g., \citealt{saunders90}). The blue long-dashed and the orange dot-dashed lines show the Schechter and modified-Schecter functions respectively, obtained by considering 
the $L_{\rm 250\mu m}$/$L_{\rm IR}$ ratio given by a spiral galaxy template at $z$$<$2 and by the IRAS~20551 template at $z$$>$2.}
\label{fig:schech}
\end{figure*}

Recent works observing sub-mm sources (e.g., SPIRE, APEX or SCUBA-2) with ALMA, report high fractions of multiple ALMA counterparts, especially for the brightest 
SPIRE 500-$\mu$m sources (e.g., \citealt{hodge13, bussmann15}), though \citet{bussmann15} find that the ALMA counterparts of the {\em Herschel} targets in some cases are located so close 
to each other that the most plausible hypothesis seems to be interactions and mergers. 
Other works observing similarly bright SPIRE 500-$\mu$m sources with SCUBA-2 seem to find very little evidence of source confusion (e.g., \citealt{bakx18}), while \citealt{hill18}
estimate the probability that a 10 mJy single-dish sub-mm source resolves into two or more galaxies to be $<$15\%.
However, the {\em Herschel} targets so far observed with ALMA are sources selected at 500-$\mu$m in SPIRE images (i.e., at longer wavelength, 
with much larger FWHM than PACS), while the {\em Herschel} catalogues considered for deriving the total IR LF are selected at PACS wavelengths (i.e., 160 $\mu$m), from maps with FWHM  
smaller than or similar to the SCUBA-2 ones at 850 $\mu$m. 
In particular, the works of \citet{gruppioni13} is based on blind 160-$\mu$m PACS catalogues, while that of \citet{magnelli13} on a PACS catalogue at 160 $\mu$m 
obtained with 24-$\mu$m prior positions. 
PACS is diffraction-limited and the photometer PSF at 160 $\mu$m is $\simeq$11 arcsec (see \citealt{poglitsch10}). 
The {\em Herschel}-PACS confusion at 100 and 160 $\mu$m is 0.15 and 0.68 mJy 
respectively (e.g., \citealt{magnelli13}), while the estimated SCUBA-2 confusion at 850 $\mu$m is $\sim$2 mJy (e.g., \citealt{chen13a}), 
therefore both PEP and SC2LS surveys are not expected to be confused at the reached fluxes.
Instead, the SPIRE confusion estimated by \citet{nguyen10} is $\sim$5.8, 6.3, 6.8 mJy at 250, 350 and 500 $\mu$m respectively, with HerMES reaching fluxes close to (or fainter than) these values.
However, the HerMES SPIRE fluxes used to construct the SEDs of the 160-$\mu$m sources have been extracted by starting from {\em Spitzer} MIPS 24-$\mu$m source positions 
(\citealt{roseboom10, oliver12}). 
Although the SPIRE instrument has FWHM of 18, 25 and 36 arcsec at 250, 350 and 500 $\mu$m respectively, the positions of sources detected at shorter wavelengths, i.e., 24 $\mu$m, 
have been utilised in order to disentangle the various contributions from discrete sources to the SPIRE flux (\citealt{roseboom10,elbaz11}). 
Thus SPIRE fluxes have been de-blended using 24-$\mu$m priors, assuming that the positions of all sources contributing significantly to the SPIRE map are known (i.e., previously detected at 24 $\mu$m), 
and that only the SPIRE flux density of each of these sources is unknown. This significantly reduces the confusion (e.g., by a factor of $\simeq$20--30 per cent, \citealt{roseboom10}),
although it might produce incomplete catalogues (i.e., missing SPIRE sources undetected in the prior band).

\subsection{The total IR LF}
\label{totIRLF}
A large discrepancy between the total IR LF derived from JCMT/SCUBA-2 and {\em Herschel} data has been ascribed to $L_{\rm IR}$ 
overestimated by {\em Herschel} works due to large {\em Herschel} beam size and source blending. However, as shown in the previous
sections, the 8--1000-$\mu$m luminosity derivation is a delicate task, depending on several factors that, if not taken into account correctly, may lead
to severe under-/over-estimation of the proper value of $L_{\rm IR}$. We believe that in the SCUBA-2 results these factors have not been considered
in the appropriate way for a fair comparison.

We note that the claimed inconsistencies between the far-IR and the sub-mm derived LFs come from very few sub-mm luminosity bins, 
likely close to the LF knee (e.g., $L^*$). In fact, if we compare Figure 3 and Figures 7 
and 8 of \citet{koprowski17}, we note that the best-fit Schechter function is obtained at 250 $\mu$m in 4 large $z$ bins from 0.5 to 4.5,
with $\geq$5 luminosity bins around $L^*$, then extrapolated to total IR luminosity and reported in all the redshift bins where 
the {\em Herschel} LF was derived (i.e., 5 $z$-bins between 0.4 and 2.3 when comparing with \citealt{magnelli13} and 10 $z$-bins between 
0.1 and 4.2 when comparing with \citealt{gruppioni13}). S2CLS data are not shown in Figure 7 and 8 of \citet{koprowski17}, but only the
Schechter function, which by-definition has a steeper bright-end than the modified-Schechter (commonly used to reproduce IR data; see \citealt{saunders90}).
The functional shape nevertheless doesn't seem to play a crucial role in the total IR LFs discrepancy, as we have verified by fitting the rest-frame 250-$\mu$m LF reported 
by \citet{koprowski17} with a modified-Schechter function, then converting it to total IR LF through the $L_{\rm 250\mu m}$/$L_{\rm IR}$ ratio given by the 
\citet{michalowski10} template, and finally interpolating in the parameter-space to match the same redshift bins of \citet{gruppioni13} 
(in order to follow exactly the same procedure described in the SCUBA-2 work).
The results of this test are illustrated in Figure~\ref{fig:schech}: the deep-pink solid line shows the best-fit modified-Schechter function and the dark-green dashed line the 
\citet{koprowski17} Schechter curve (both converted from 250-$\mu$m, as described above). We find that the two functions are not significantly different in the plotted range,
in any case not enough to explain the discrepancy with data, although both curves obtained from our test appear closer to the {\em Herschel}-PEP data points (black dots) than 
the one reported in Figure~8 of \citet{koprowski17}. The bright-end of the total IR LFs at $z$$>$1 is not well reproduced by either the Schechter nor the modified-Schechter 
function, although a much better agreement with the data is obtained by considering the $L_{\rm 250\mu m}$/$L_{\rm IR}$ ratio for a spiral galaxy template,
and for the IRAS~20551 template, respectively at $z$$<$ and $>$2
(see, e.g., the orange dot-dashed and the blue dot-dot-dot-dashed lines in Figure~\ref{fig:schech}, showing the modified-Schechter and the Schechter function respectively, obtained with these ratios). 
These latter templates are more similar to the average SEDs of the populations that dominate the {\em Herschel}-PEP LFs in the two redshift intervals (although the calculation 
reported here is rougher than a proper one performed on an object-by-object basis). 
From this test we can therefore conclude that the considered template SEDs, combined with incompleteness issues, are likely the principal players in the observed discrepancies.

\subsection{The 850-$\mu$m Source Counts}
\label{n850mic}
\begin{figure}
	\includegraphics[width=8cm,height=7.5cm]{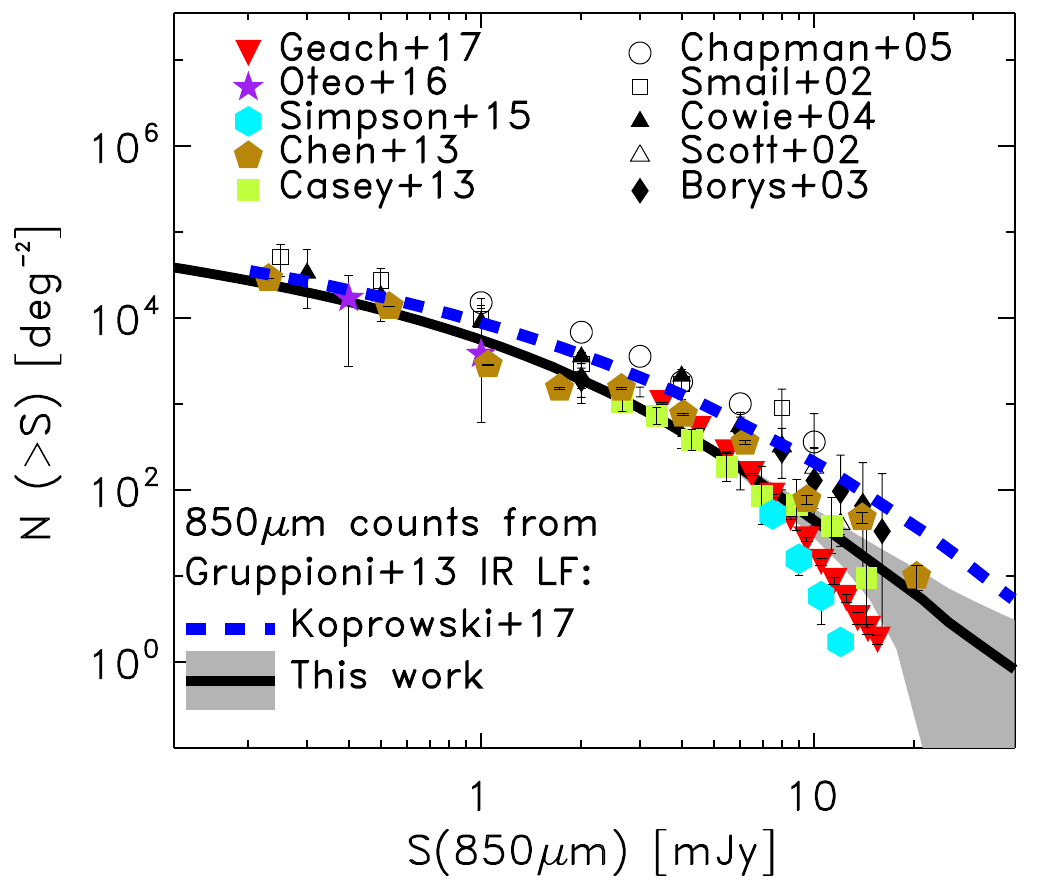}
    \caption{Integral source counts at 850 $\mu$m from 
SCUBA, SCUBA-2 and ALMA data from the literature are shown as different symbols (as explained in the legend) and compared with the number counts 
 predicted from the {\em Herschel} total IR LF (\citealt{gruppioni13}), as derived in this work, shown by the black solid line within grey uncertainty area, and
 by \citet{koprowski17}, shown by the blue dashed line.}
\label{fig:850um_cnt}
\end{figure}
The comparison between the 850-$\mu$m source counts obtained from the S2CLS and from other literature surveys, and the ones estimated from the 
{\em Herschel} total IR LFs by \citet{koprowski17} also showed a significant inconsistency (see Figure~9 of \citealt{koprowski17}). In particular, 
the {\em Herschel}-derived 850-$\mu$m counts seemed to predict an order of magnitude more sources than observed at bright 850-$\mu$m flux
densities (e.g., $>$10 mJy). This severe over-prediction    
of the 850-$\mu$m counts produced by the {\em Herschel} IR LFs led the authors to the conclusion
that the high values reported from these studies most likely reflect problems in source identification 
and redshift estimation arising from the large-beam long-wavelength SPIRE data, as well as
potential blending issues.
However, to obtain this result \citet{koprowski17} used a single SED for all the {\em Herschel} sources and a 
unique evolution for the whole LF (while different evolutionary paths had been found 
for different populations; \citealt{gruppioni13}).
Here we demonstrate how the simple use of different ingredients (e.g., the best-fit SED for each source and different evolutions for the different IR populations) 
provides results based on the {\em Herschel} IR LF much closer to the observations, though without any fine tuning aiming at reproducing the 
observed 850-$\mu$m source counts and redshift distributions. 
In Figure~\ref{fig:850um_cnt} we show the 850-$\mu$m integral source counts
obtained by integrating the {\em Herschel} LFs and the
evolutions found for each SED-class of {\em Herschel} galaxies by \citet{gruppioni13}, then converting to sub-mm wavelength using the best-fit SEDs for each source. 
Note that the {\em Herschel} LF is well constrained by data only to $z$$\sim$3, with an upper limit derivation at 3.0$<$$z$$\leq$4.2.
At higher redshifts we need to extrapolate the evolutions: the results shown in the plot have been obtained by assuming a negative density evolution for all the populations 
(e.g., $\propto$(1$+$$z$)$^{-3}$) at $z$$>$3. The bright-end of the 850-$\mu$m counts is slightly overestimated if we compare to the \citet{geach17} and \citet{simpson15}
works, but is very close to the \citet{casey13} and \citet{chen13b} points. In any case, it is very far (and much closer to the data) from what
\citet{koprowski17} claim to be the prediction obtained by integrating the \citet{gruppioni13} evolving LFs. 
At fluxes $<$10 mJy the agreement between our {\em Herschel} predictions and the sub-mm observations is very good. 

\section{Conclusions}
\label{concl}
In order to understand the cause of the large discrepancies observed in the total IR LFs derived from {\em Herschel}
and SCUBA-2 survey data, we have run through all the steps that lead to the total IR LF calculation, by analysing in detail all 
the assumptions that can be made and the incompleteness than can affect the selection at different wavelengths (i.e., 160 $\mu$m or 850 $\mu$m).
We have shown how the choice of different ingredients can lead to estimates of
$L_{\rm IR}$ and, consequently, of the total IR LF, that can significantly disagree with each other. 

The main conclusions of this work can be summarised as follows:
\begin{itemize}
\item Since a wide diversity of SEDs is found in far-IR selected samples of galaxies,
and since different SED-types are found to dominate the IR populations at different redshifts, the use of a single template for a whole survey
is probably too restrictive: the more data are considered for constraining the galaxy SEDs (especially around the far-IR bump), the more 
precisely $L_{\rm IR}$ is derived. If the luminosities are inaccurate, objects are counted in the wrong luminosity bin, biasing the resulting
total IR LF. 
\item If the sample completeness is not accurately estimated and corrected, the resulting source number density can be wrong.
In particular, the current SCUBA-2 surveys appear to be incomplete (mainly at $z$$<$3) against galaxies with ``warm'' SEDs,
which are indeed found to be the major contributors to the bright-end of the {\em Herschel} IR LF and of the SFRD at $z$$\gsimeq$2.  
\item The functional form considered to fit the LF data (i.e., Schechter or modified-Schechter function) does not seem to play a major role   
in the total IR LFs discrepancies. 
\item By integrating the {\em Herschel} total IR LFs, considering the best-fit SED for each source and different evolutions for each galaxy
population (as derived by \citealt{gruppioni13}), we have obtained an estimate of the 850-$\mu$m source counts in agreement with
observations (and different from what has been obtained by \citet{koprowski17} by integrating the same IR LFs).
\end{itemize}

We therefore conclude that the observed differences have to be ascribed principally to the use of a single SED template in the calculations, 
and to a considerable incompleteness against a significant population of IR galaxies detected in far-IR surveys, but not in sub-mm ones. 

Finally, we must note that the {\em Herschel} PACS LFs (i.e., \citealt{gruppioni13}, \citealt{magnelli13}) are in perfect agreement with totally independent derivations,
either from $Spitzer$-24 $\mu$m (\citealt{rodighiero10}), SPIRE (\citealt{lapi11,rowanrobinson16,marchetti16}), VLA-3 GHz  (\citealt{novak17}) and -35 GHz data (\citealt{riechers18}).
The latter work, reporting the measurement of the CO luminosity function at $z$$\sim$2--3 and $\sim$5--7, as part of the CO Luminosity Density at High Redshift (COLDz) survey,
shows a very good agreement with the empirical predictions by \citet{vallini16}, which are indeed based on the \citet{gruppioni13} {\em Herschel} IR LF. In fact, the
\citet{vallini16} CO LF seem to be the only ones reproducing the excess of bright CO sources compared to the semi-analytical predictions observed at high-$z$.  
Moreover, the new method to super-deblend {\em Herschel} images based on priors at other wavelenghts presented by \citet{liu17}, provides values of
SFRD very similar to those previously derived by \citet{gruppioni13} and \citet{magnelli13}.


\section*{Acknowledgements}
We thank an anonymous referee and the scientific editor for helpful comments that led to 
significant improvements in the paper. 
A particular thank goes to M. Negrello and G. De Zotti for fruitful discussions and
for urging us to work on this subject.
We acknowledge funding from the INAF PRIN-SKA 2017 program 1.05.01.88.04. 




\bibliographystyle{mnras}
\bibliography{mybibliography}


\bsp	
\label{lastpage}
\end{document}